\documentclass[aps,prb,10pt,twocolumn]{revtex4-1}
\usepackage{amsmath}
\usepackage{amsfonts}
\usepackage{amssymb}
\usepackage{graphicx}
\usepackage[colorlinks=true,linktocpage=true,breaklinks=true]{hyperref}

\begin{document}

\title{Spin-orbit torque in two dimensional antiferromagnetic topological insulators}
\author{S. Ghosh}
\email{sumit.ghosh@kaust.edu.sa}
\author{A. Manchon}
\email{aurelien.manchon@kaust.edu.sa}
\affiliation{King Abdullah University of Science and Technology (KAUST), Physical Science and Engineering Division (PSE), Thuwal 23955, Saudi Arabia}

\begin{abstract}
We investigate spin transport in two dimensional ferromagnetic (FTI) and antiferromagnetic (AFTI) topological insulators. In presence of an in plane magnetization AFTI supports zero energy modes, which enables topologically protected edge conduction at low energy. We address the nature of current-driven spin torque in these structures and study the impact of spin-independent disorder. Interestingly, upon strong disorder the spin torque develops an antidamping component (i.e. {\em even} upon magnetization reversal) along the edges, which could enable current-driven manipulation of the antiferromagnetic order parameter. This antidamping torque decreases when increasing the system size and when the system enters the trivial insulator regime. 
\end{abstract}

\maketitle

\section{Introduction}
The successful manipulation of small magnetic elements using spin-polarized currents via spin transfer torque has opened appealing perspectives for low power spin devices \cite{Slonczewski1996, Berger1996, Chappert2007}. In the past ten years, it has been predicted \cite{Bernevig2005, Manchon2008, Obata2008,Garate2009} and observed \cite{Chernyshov2009, Endo2010,Miron2010,Liu2011,Miron2011,Liu2012} that noncentrosymmetric magnets with large spin-orbit coupling can also exhibit
large spin torque, a phenomenon called spin-orbit torques (SOT). The physics of SOT in homogeneous ferromagnets
\cite{Matos-Abiague2009, Hals2010, Pesin2012, Bijl2012, Wang2012a, Qaiumzadeh2015, Li2015} and magnetic textures \cite{Kononov2014, Stier2014, Linder2013, Kim2012, Khvalkovskiy2013, Thiaville2012} has attracted a massive amount of attention since then. While these torques have been originally studied in bulk non-centrosymmetric magnets \cite{Chernyshov2009,Endo2010} and ultrathin magnetic multilayers \cite{Miron2010, Liu2011, Miron2011, Liu2012}, their observation has been recently extended to magnetic bilayers involving topological insulators \cite{Mellnik2014, Fan2014, Wang2015, Fan2016}.

A topological insulator (TI) is characterized by gapless edge/surface states in the absence of external magnetic field \cite{Qi2011}. The zero energy modes arise due to time reversal symmetry and are immune to nonmagnetic disorder \cite{Koenig2007, Wang2012}. This topological protection breaks down in presence of magnetization which destroys the zero energy modes and opens a gap \cite{Liu2009}. This process is accompanied by the emergence of quantum anomalous Hall effect \cite{Liu2008, Chang2013, Chang2015}, as well as quantum magnetoelectric effect when the Fermi level lies in the gap of the surface states \cite{Qi2008, Garate2010}. Recently, three dimensional TI have been used to achieve large SOT in an adjacent ferromagnet \cite{Mellnik2014, Fan2014, Fan2016, Wang2015}. In spite of significant theoretical efforts to model the SOT exerted on homogeneous ferromagnets \cite{Taguchi2015, Fujimoto2014, Sakai2014, Mahfouzi2014, Chang2015, Mahfouzi2016, Ndiaye,Fischer2016} and magnetic textures \cite{Yokoyama2010, Yokoyama2010a, Nomura2010, Yokoyama2011, Tserkovnyak2012, Tserkovnyak2015}, the exact nature of the  torque observed experimentally remains a matter of debate as it is not clear whether surface states are still present, and how bulk and surface transport contribute to the different components of the torque. Besides significant challenges in terms of materials growth, the main difficulty lies in the fact that magnetism itself breaks the topological protection of surface states \cite{Chen2010, Wray2011, Checkelsky2012, Luo2013, Eremeev2013,Zhang2016}, which prevents from taking full advantage of the gigantic spin-orbit coupling of the Dirac cones. Fortunately, ferromagnetism is not the only useful magnetic order parameter that appears in nature.

 Recently, it has been realized that antiferromagnets can also be controlled by spin transfer torque \cite{Nunez2006,Gomonay2010,MacDonald2011}, opening the emergent field of antiferromagnetic spintronics \cite{Gomonay2014, Jungwirth2016, Baltz}. The nature of spin transfer torque has been investigated theoretically in antiferromagnetic spin-valves and tunnel junctions
 \cite{Duine2007, Haney2007, Haney2008, Xu2008, Haney2008a, Prakhya2014, Merodio2014, Saidaoui2014}, as well as antiferromagnetic domain walls \cite{Hals2011, Swaving2011, Tveten2013, Gomonay2016, Shiino2016, Selzer2016, Yamane2016}. Most importantly for the present work, it has been recently predicted \cite{Zelezny2014, Zelezny} and experimentally demonstrated \cite{Wadley2016} that SOT can also be used to control the direction of the antiferromagnetic order parameter. This naturally brings TI as a possible testing ground due to their inherent strong spin-orbit coupling. Since antiferromagnetism only breaks time-reversal symmetry locally but not globally, it preserves the topological nature of the surface gapless states \cite{Mong2010,Baireuther2014}. Exploring the possibility of combining the topological nature of the surface or edge states in antiferromagnetic topological insulators with the physics of SOT could therefore open appealing perspectives.

In this work, using scattering wave function formalism implemented on a tight-binding model, we explore the nature of spin transport and torque in two dimensional ferromagnetic (FTI) and antiferromagnetic (AFTI) topological insulators. We find that AFTI is more robust against disorder than FTI, such that topological edge states are preserved even under weak disorder. Most importantly, SOT possesses two components: a field-like torque ({\em odd} under magnetization reversal) and an antidamping torque ({\em even} under magnetization reversal). While the former is directly generated by the spin-momentum locking at the edges, the latter arises upon scattering and is quite sensitive to disorder and size effects.

\section{Method}

We start from the Bernevig-Hughes-Zhang model \cite{Bernevig2006} on a square lattice. We use the basis $(1\uparrow,2\uparrow,1\downarrow,2\downarrow)^T$, where $1,2$ refer to two orbitals and $\uparrow,\downarrow$ refer to spin projections, and define the TI Hamiltonian by a $4\times4$ matrix,

\begin{eqnarray}
H(k) = \left(
\begin{array}{cc}
h(k) & 0 \\
0 & h^*(-k)
\end{array}
\right),
\label{BHZ}
\end{eqnarray}

where $h(k)$ is given by

\begin{eqnarray}
h(k) = & [M + B (\cos(k_x) + \cos(k_y))]\sigma_z \nonumber \\
 & + A \sin(k_x) \sigma_x + A \sin(k_y) \sigma_y.
\end{eqnarray}
Here $A,B,M$ are model parameters whose values depend on the real structure \cite{Bernevig2006,Qi2011}. The topologically nontrivial phases appear for $B>|M/2|$, which is manifested as gapless edge states in quasi one dimensional systems. In case of a CdTe-HgTe quantum well this is achieved by tuning the width of the quantum well. For our calculations we choose $A$ to be the unit of energy and consider $B=1.0A, M=-1.5A$ that ensures the existence of topologically protected edge states for nonmagnetic TI. To map this bulk Hamiltonian  (\ref{BHZ}) on a finite scattering region, we first extract the tight-binding parameters \cite{Mong2010,Dang2014} by expressing

\begin{eqnarray}
H(k) = H_{0}+ H_{\hat{x}}e^{i k_x}+H_{\hat{y}}e^{i k_y} + H_{\hat{x}}^\dagger e^{-i k_x}+H_{\hat{y}}^\dagger e^{-i k_y}, \nonumber \\
\label{h1}
\end{eqnarray}
with
\begin{eqnarray}
H_{0} &=& \left(
\begin{array}{cccc}
 M & 0 & 0 & 0 \\
 0 & -M & 0 & 0 \\
 0 & 0 & M & 0 \\
 0 & 0 & 0 & -M 
\end{array}
\right), \nonumber \\
H_{\hat{x}} &=& \frac{1}{2}\left(
\begin{array}{cccc}
B & -iA & 0 & 0 \\
-iA & -B & 0 & 0 \\
0 & 0 & B & iA \\
0 & 0 & iA & -B
\end{array}
\right),\nonumber \\
H_{\hat{y}} &=& \frac{1}{2}\left(
\begin{array}{cccc}
B & -A & 0 & 0 \\
A & -B & 0 & 0 \\
0 & 0 & B & -A \\
0 & 0 & A & -B
\end{array}
\right).\nonumber\end{eqnarray}
We can use these hopping elements to construct a real space Hamiltonian for a finite system as 

\begin{eqnarray}
H &=& \sum_{i} c_i^\dagger H_0 c_i + \sum_{i \neq j} c_i^\dagger t_{ij}c_j\label{hreal}
\end{eqnarray}
where $t_{ij}=H_{\hat{r}_{ij}}$, $\hat{r}_{ij}$ (=$\pm\hat{x}$, $\pm\hat{y}$) being the unit vector between nearest neighbor sites $i$ and $j$, and $c_i^\dagger(c_i)$ is the creation (destruction) operator for the state $(1\uparrow,2\uparrow,1\downarrow,2\downarrow)^T$ at $i$-th site. The coupling between itinerant spins $\vec{s}$ and the local magnetization ($\vec{m}_i$), as well as the disorder potential $V_i^{r}$ are introduced in the onsite energy as
\begin{eqnarray}
H_m^{imp} = \sum_{i}  c_i^\dagger (H_0 + \vec{m}_i \cdot\vec{s} + V_i^{r} \mathbb{I}_4) c_i + \sum_{i \neq j}c_i^\dagger t_{ij} c_j,
\label{Himp} 
\end{eqnarray}
where $\mathbb{I}_n$ is the $n$-th rank identity matrix, and $\vec{s}=(\hat{s}_x,\hat{s}_y,\hat{s}_z)$, with
\begin{eqnarray}
& \hat{s}_x=\left(
\begin{array}{cc}
0 & \mathbb{I}_2 \\
\mathbb{I}_2 & 0
\end{array}
\right), 
\hat{s}_y=\left(
\begin{array}{cc}
0 & -i\mathbb{I}_2 \\
i\mathbb{I}_2 & 0
\end{array}
\right),   
\hat{s}_z=\left(
\begin{array}{cc}
\mathbb{I}_2 & 0 \\
0 & -\mathbb{I}_2
\end{array}
\right). \nonumber \\   
\label{mv}
\end{eqnarray}

\begin{figure}[h]
\centering
\includegraphics[width=0.35\textwidth]{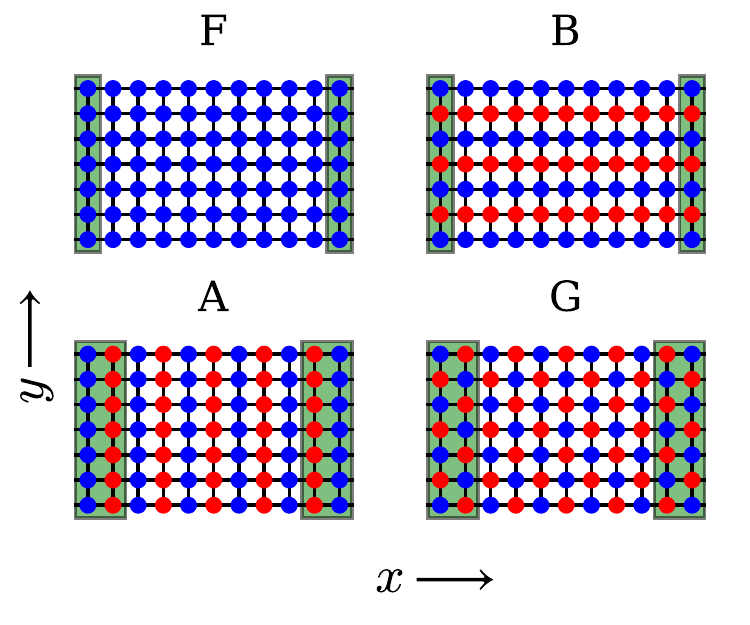}
\caption{Schematic of FTI and different types of AFTI. The green region shows one unit cell of the lead. Blue and red dots represent positive and negative  $\vec{m}_i$.}
\label{afs0}
\end{figure}

In the following, we consider five different configurations, defined by the spatial modulation of $\vec{m}_i$: an ordinary, non-magnetic TI as a reference (referred to as $O$), an FTI ($F$), and A-, B-, and G-type AFTI configurations [($A$), ($B$) and ($G$) in Fig.~\ref{afs0}]. The total system can be divided into three parts - (i) left lead, (ii) scattering region and (iii) right lead. The leads are semi-infinite and can be characterized by one unit cell (green shaded region in Fig.~\ref{afs0}) whereas the scattering region is defined by Eq. \eqref{Himp}. Note that for $A,G$ type AFTI we need to double the unit cell of the lead to maintain the translation symmetry. For this work, we consider a scattering region composed of $40\times 20$ sites arranged on a square lattice. To calculate the transport properties we adopt the wavefunction approach, as implemented in the tight-binding software KWANT \cite{Groth2014}. This approach is equivalent to the non-equilibrium Green's function formalism \cite{Fisher1981}.
In this method one starts by defining the incoming modes at a particular energy in terms of eigenstates of an infinite lead and subsequently obtain the wavefunction within the scattering region by using the continuity relations. By applying this method throughout the scattering region one can obtain the outgoing modes that can be exploited to construct the $S$-matrix of the system. The scattering wavefunction and the $S$-matrix are two basic outputs one can obtain from KWANT for any given system (see Section 2 of Ref. \onlinecite{Groth2014} for details).  
The conductance of the system is calculated from the $S$-matrix using  Landauer-B\"uttiker formalism.  
To calculate the non-equilibrium spin density at some given energy $E_F$, we use a small bias voltage $V_{Bias}=\mu_L-\mu_R$, where $e\mu_{L(R)}=E_F \pm eV_{Bias}/2$ are the chemical potential of the left (right) lead. We use the scattering wavefunction calculated by KWANT and evaluate the expectation values of different spin components integrated over the bias window to get the total non-equilibrium spin density as,
\begin{eqnarray}
\vec{S}_i^{neq} = \int_{\mu_R}^{\mu_L} \langle \psi_i(E) \vert \vec{s} \vert \psi_i(E) \rangle dE,
\label{neq}
\end{eqnarray}
where $\vec{s}$ is the onsite spin operator defined in Eq. \eqref{mv}, and $\psi_i(E)$ is the scattering wavefunction for $i$-$th$ site at energy $E$. Once we get the non-equilibrium spin density we can calculate the onsite SOT as
\begin{eqnarray}
\vec{\tau_i} = \vec{m_i} \times \vec{S}_i^{neq}.
\label{SOT} 
\end{eqnarray}

Finally, in order to introduce nonmagnetic disorder in the system we add to the Hamiltonian, Eq. \eqref{Himp}, a random onsite energy $V_{i}^r$ uniformly distributed over the range [$-V_0, V_0$]. This gets rid of any possible shift of energy spectrum that might appear if one chose only positive amplitudes for the disorder potential. The transport properties are then averaged over 1280 random disorder configurations.

\section{Robustness of different magnetic configurations}

\begin{figure}[h]
\centering
\includegraphics[width=0.23\textwidth]{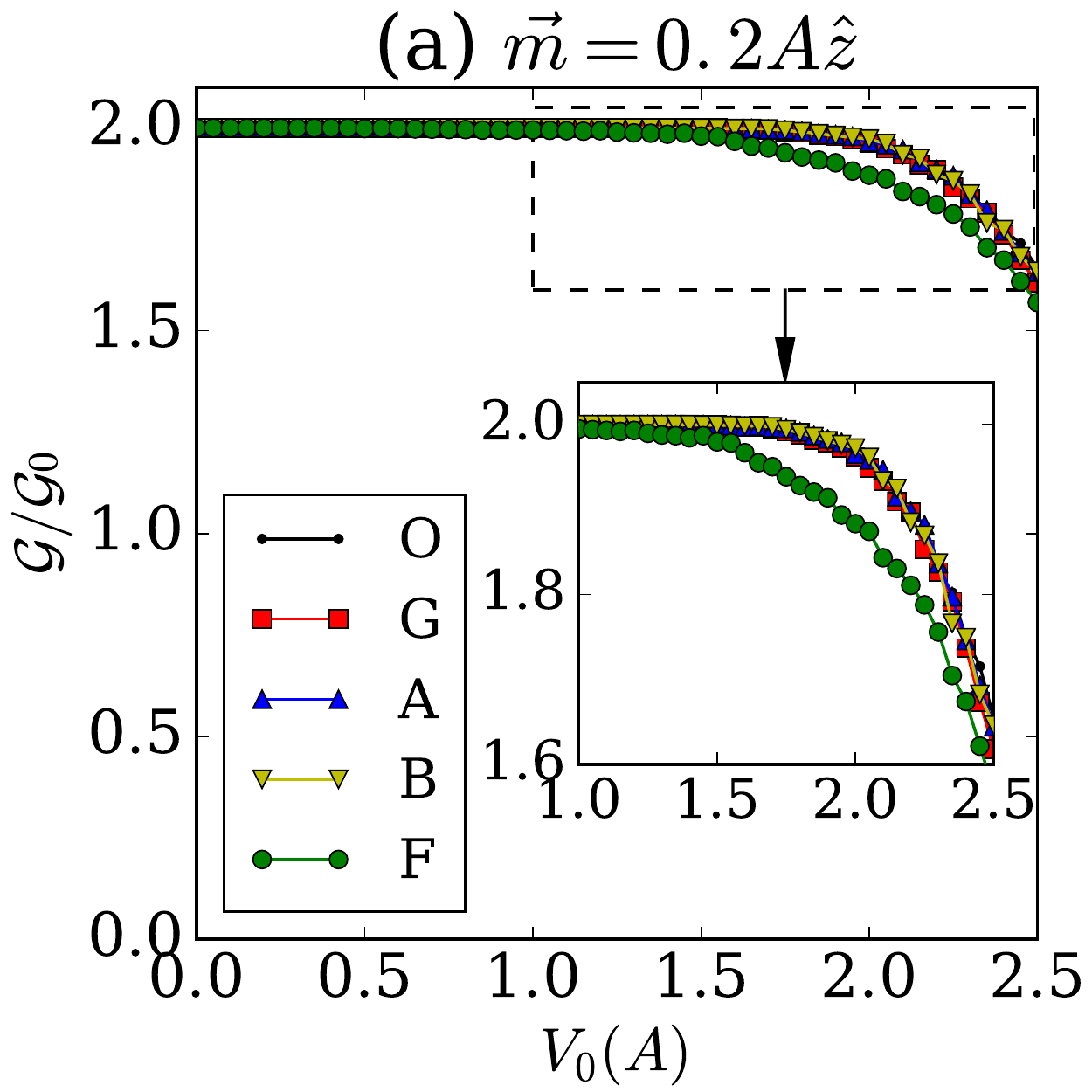}
\includegraphics[width=0.23\textwidth]{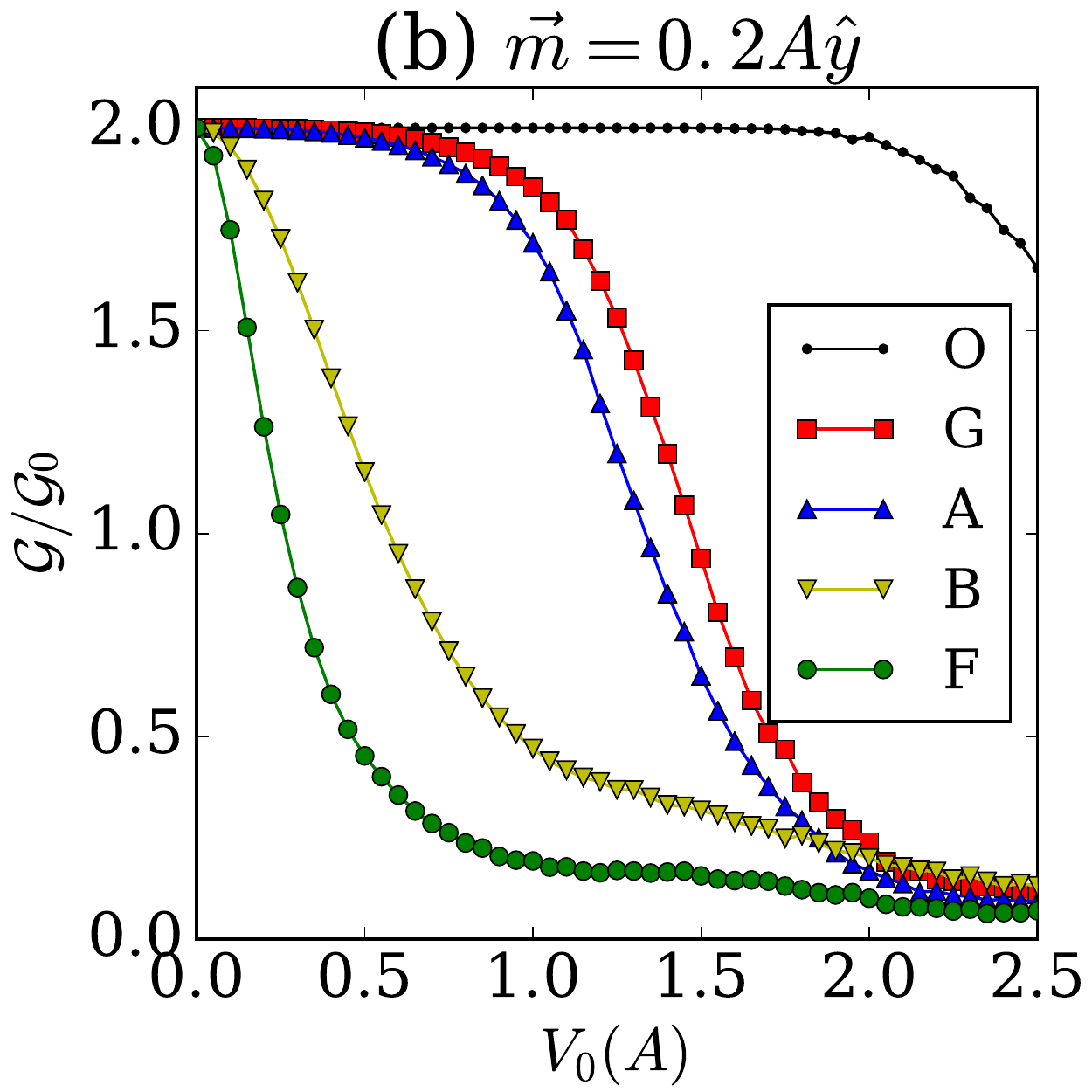}
\caption{Conductance ($\mathcal{G}$) of TI with different magnetic configurations against random disorder for (a) out of plane and (b) in plane magnetic moments. The conductance is normalized to the conductance quantum $\mathcal{G}_0=2e^2/h$. The boxed portion of (a) is enlarged in the inset.}
\label{afs}
\end{figure}

Let us first compute the impact of disorder on the conductance in the various magnetic configurations. Fig. \ref{afs}(a,b) displays the behavior of conductance as a function of the disorder strength when the direction of the magnetic moments $\vec{m}_i$ is (a) out of plane and (b) in the plane. Here, the transport energy is taken $E_F=0.25A$.

From Fig.~\ref{afs}(a) we see that when the magnetic moments lie out-of-plane FTI (F) is comparatively more sensitive to disorder than AFTI (A, B, G) and nonmagnetic TI (O), although the difference of robustness between the nonmagnetic and magnetic TI is not very large. The initial quantized conductance of FTI starts decreasing around $V_0\sim 1.5A$ due to the progressive quenching of the topological protection of the edge modes, while in contrast, AFTI  and nonmagnetic TI maintain their topological egde states up to $V_0\sim 2A$. The difference becomes quite significant when we set the magnetic order in the plane, see Fig.~\ref{afs}(b). Noticeably, F and B cases are very sensitive to disorder, while A and G cases are much more robust. This indicates that AFTIs with an in-plane staggered magnetic character along the transport direction remain topological insulators even for weak disorder. 

\begin{figure}[h]
\centering
\includegraphics[width=0.23\textwidth]{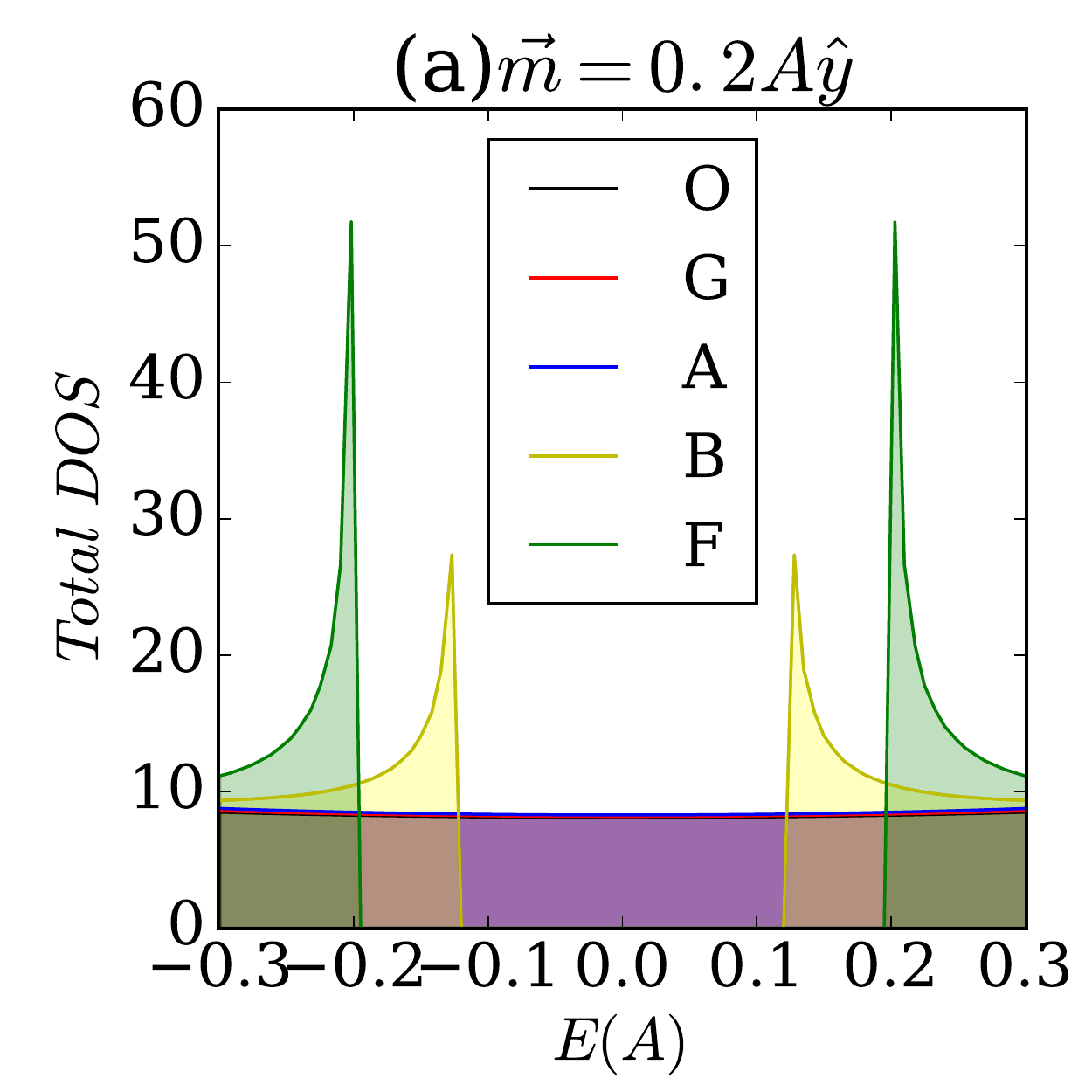}
\includegraphics[width=0.23\textwidth]{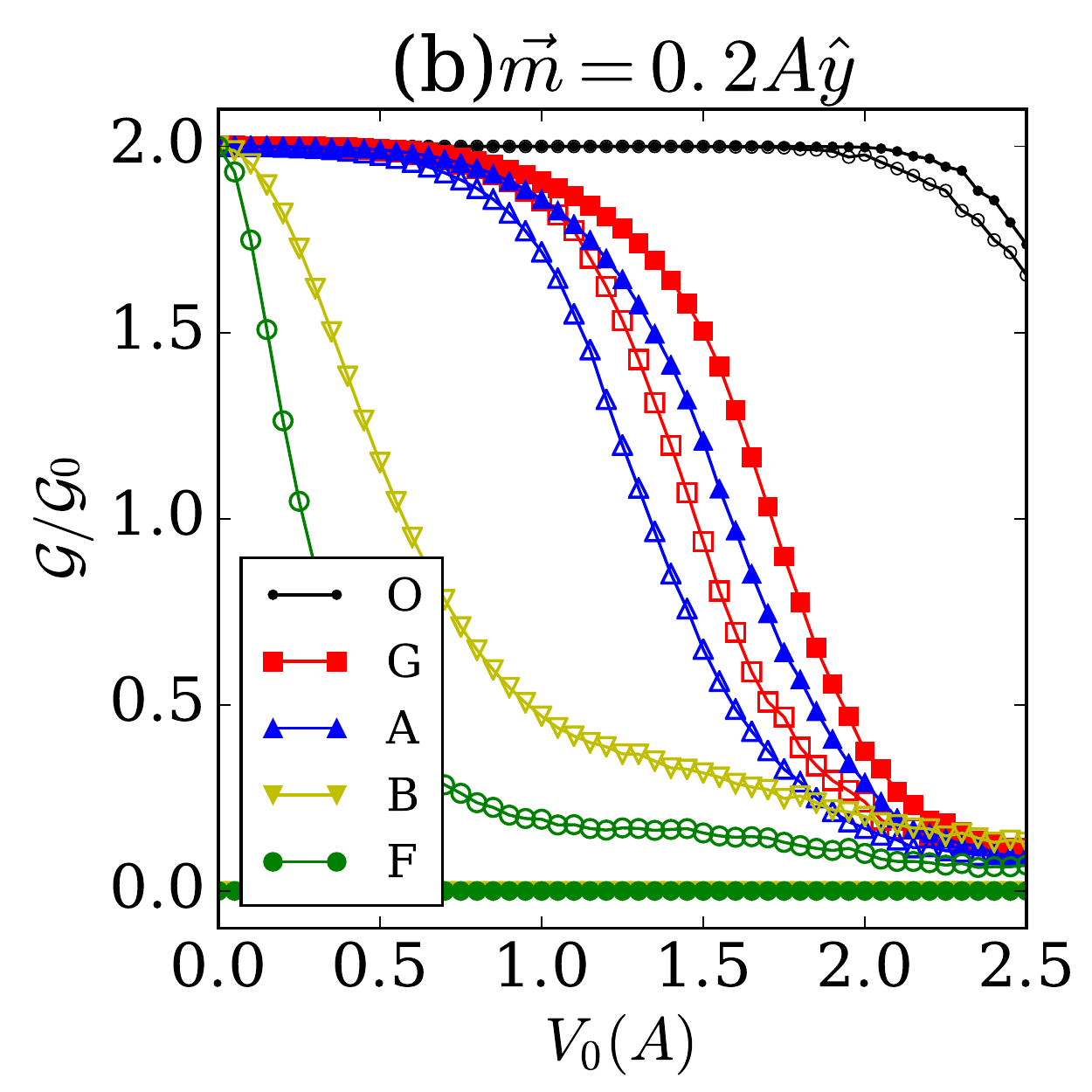}
\caption{(a) Total density of states (DOS) for TI, FTI and different AFTI. (b) Conductance against random disorder at $E_F = 0.05A$ (filled symbols) and at $E_F=0.25A$ (open symbols) for different configurations with an in-plane magnetic moment $0.2A\hat{y}$.}
\label{condef}
\end{figure}

For a better understanding of this effect, we calculate the density of states of the TIs in the absence of disorder, see Fig. \ref{condef}(a). An in-plane magnetic order opens a gap for F and B, while A and G can preserve their gapless states similarly to the nonmagnetic TI (O). Since the topological protection is stronger at lower energy, we calculate the robustness at $E_F=0.05A$ [Fig.~\ref{condef}(b)] and find the quantized conductance due to the edge states of A, G and O cases survives longer compared to that evaluated at $E_F=0.25A$. B and F types have a gap at that energy and hence show zero conductance.

From now on, we proceed with only FTI and G-type AFTI as they qualitatively behave similarly as B-type AFTI and A-type AFTI, respectively.

\section{Non-equilibrium spin density and Spin-orbit torque}
As mentioned in the introduction, spin transfer torque \cite{Nunez2006,Gomonay2010,MacDonald2011} as well as SOT \cite{Zelezny2014,Zelezny} can be used to control the direction of the antiferromagnetic order parameter. The order parameter can be controlled in two ways \cite{Gomonay2014,Jungwirth2016,Baltz}: either using a time-dependent (ac) field-like torque (i.e., a torque that is {\em odd} under magnetization reversal), or using a time-independent (dc) antidamping torque (i.e., {\em even} under magnetization reversal). Our intention is to investigate the nature of SOT in the AFTI case, where both topologically protected edge transport and antiferromagnetic order parameter coexist.\par

\begin{figure}[h]
\centering
\includegraphics[width=0.235\textwidth]{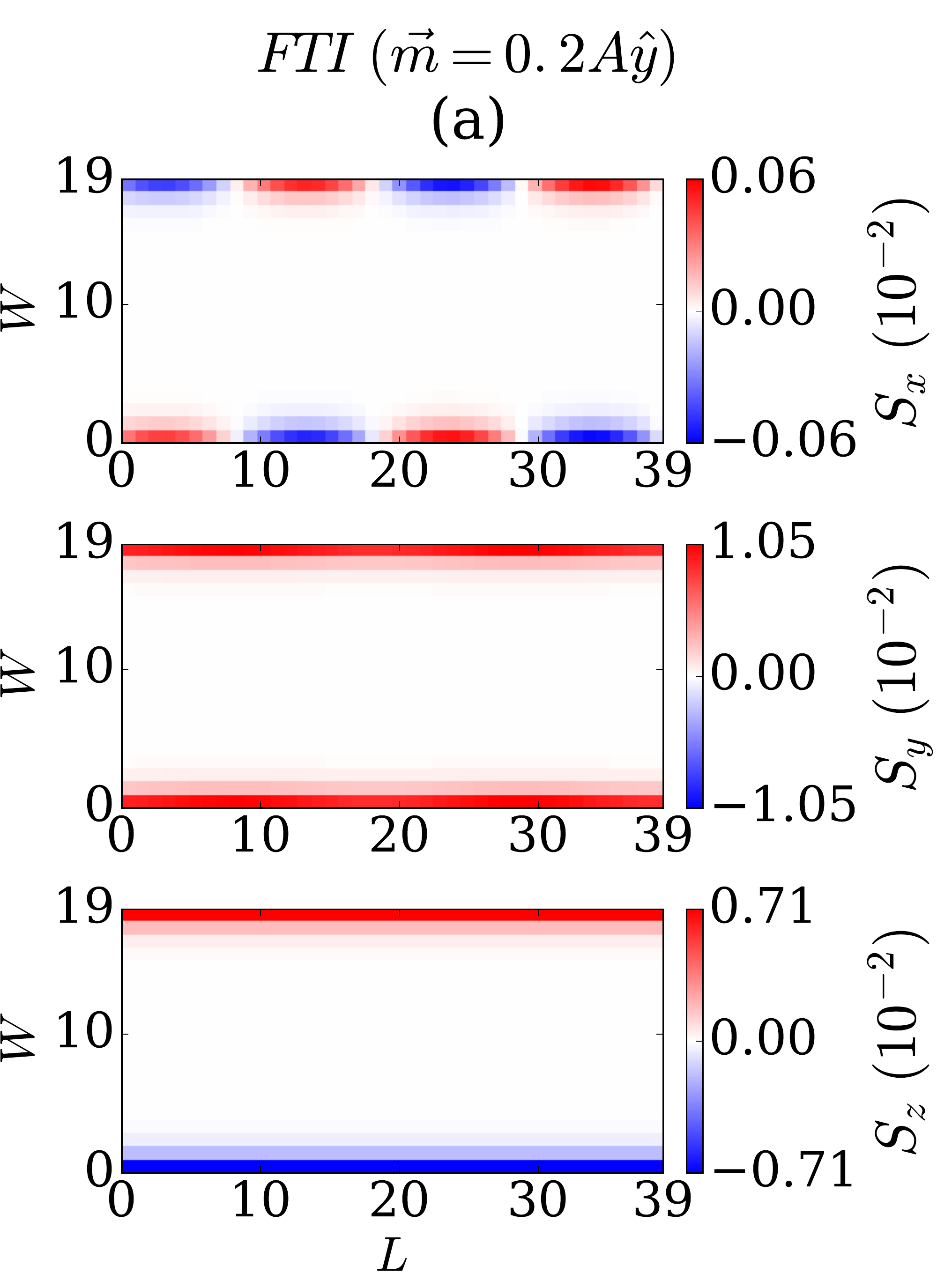}
\includegraphics[width=0.235\textwidth]{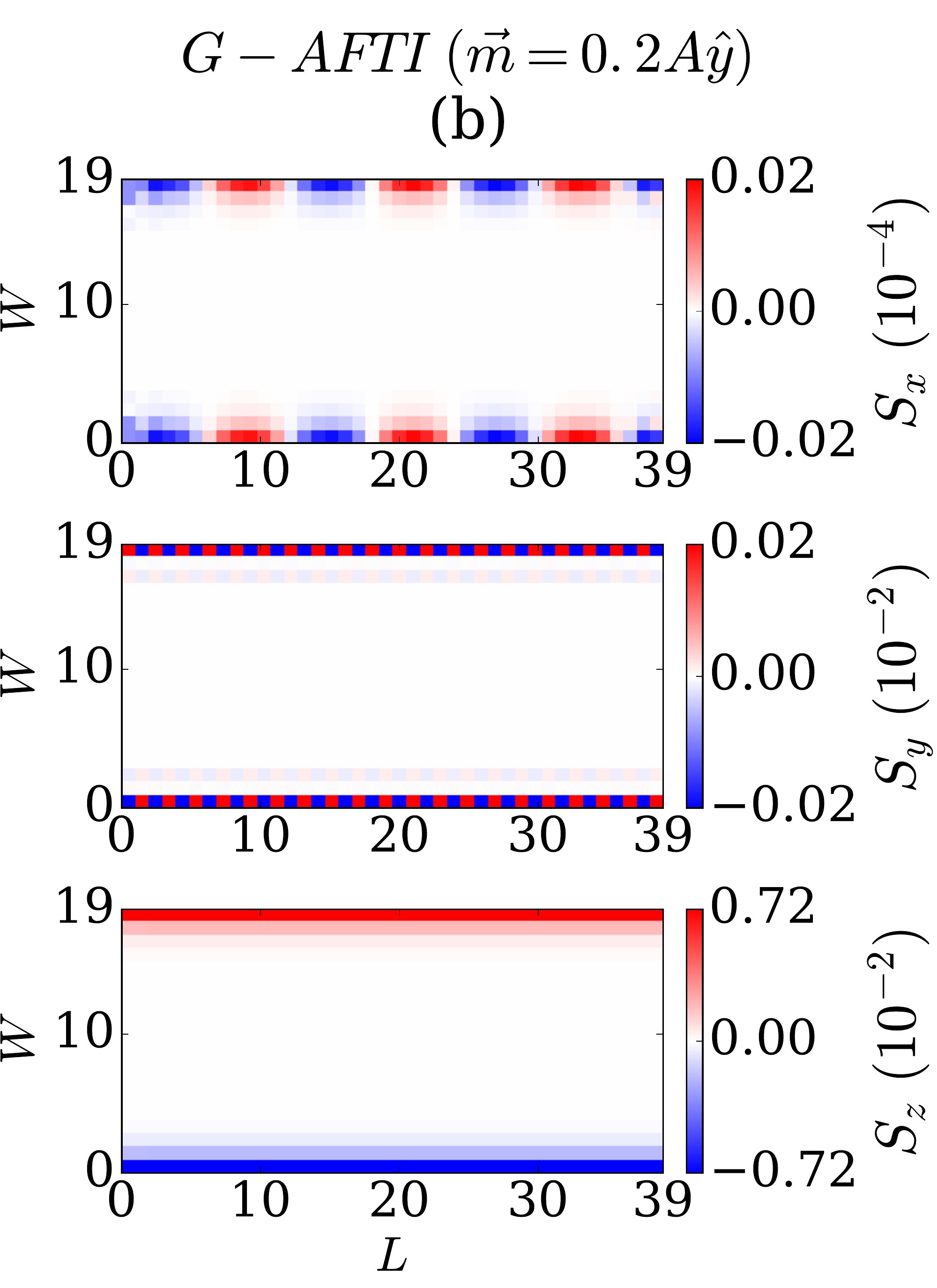} \\
\includegraphics[width=0.23\textwidth]{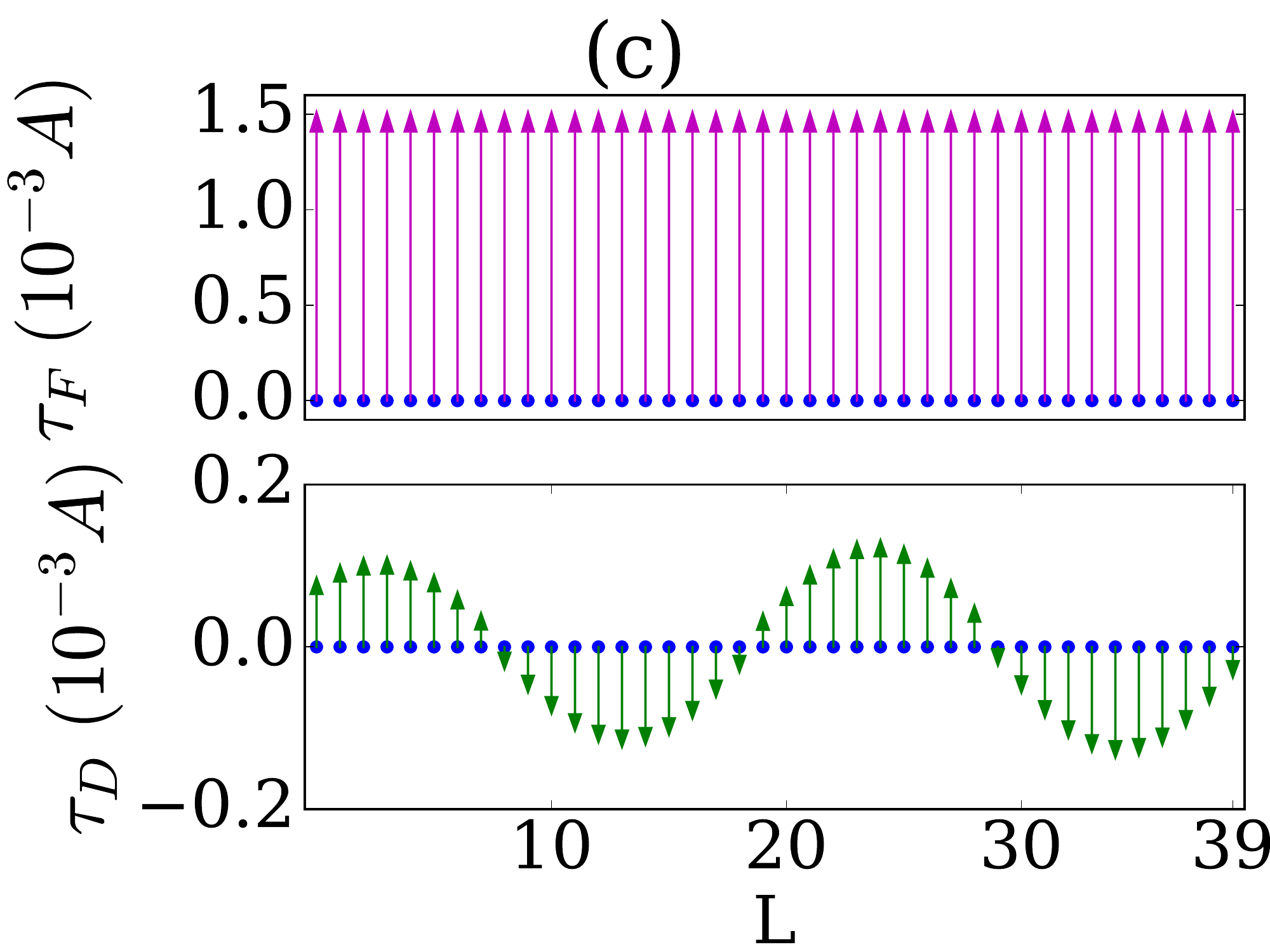}
\includegraphics[width=0.23\textwidth]{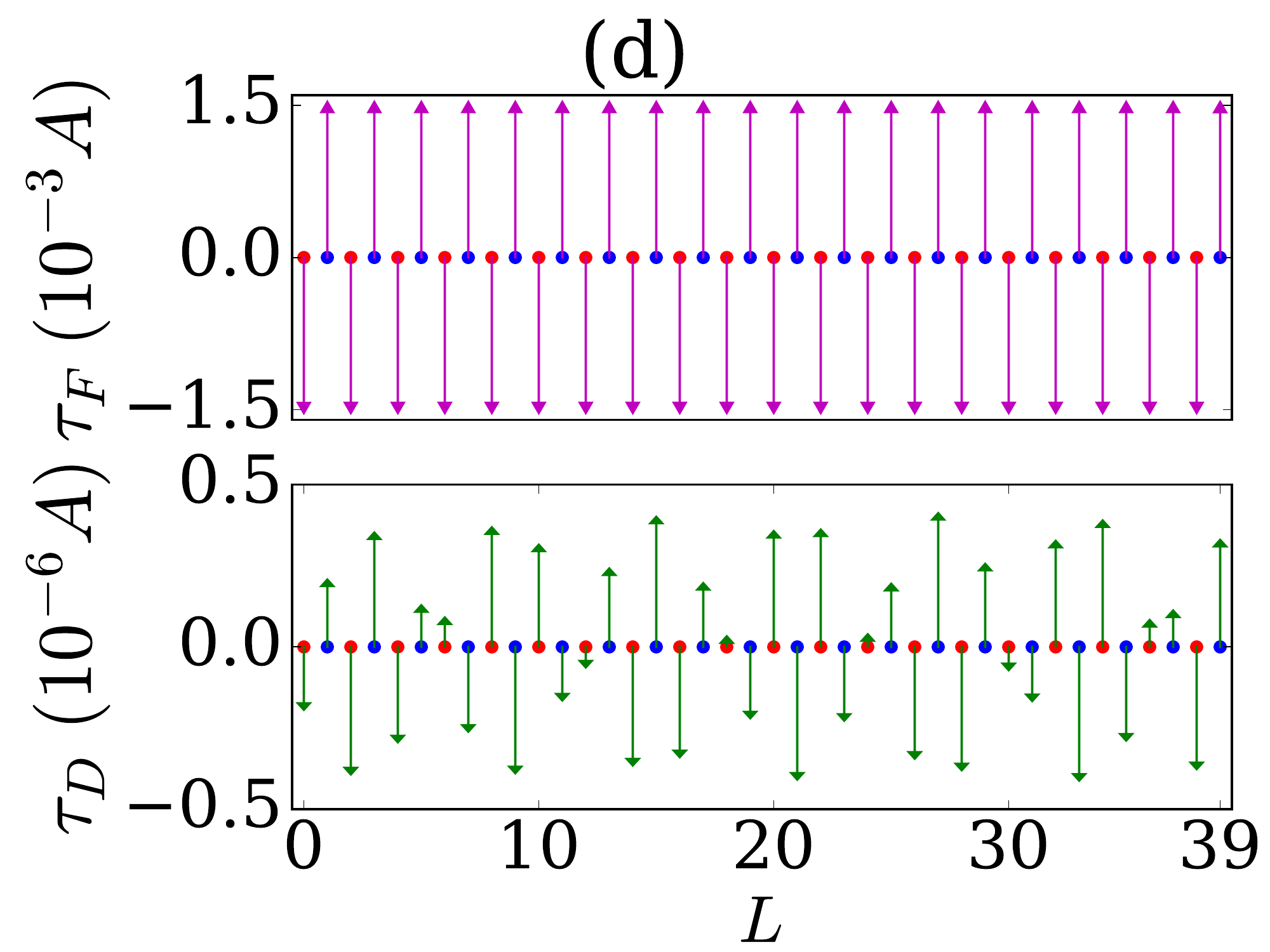}
\caption{Non-equilibrium spin density for (a) FTI and (b) G-AFTI. We use a magnetization strength $m=0.2A$ along $\hat{y}$ and the spin densities are evaluated at $E_F=0.25A$ with a bias voltage $(\mu_ L-\mu _R)=0.02A$. (c,d) shows the corresponding field like ($\tau_F$, magenta) and antidamping ($\tau_D$, green) SOT evaluated at the top edge for FTI and G-AFTI respectively.} \label{spd}
\end{figure}

First we calculate the total non-equilibrium spin density and the associated SOT in FTI [Fig.~\ref{spd}(a,c)] and AFTI [Fig.~\ref{spd}(b,d)], with an in-plane magnetic order ($\vec{m}_i\sim\hat{y}$) and in the absence of disorder. In these calculations, we set $|m_i|=0.2A$, $E_F=0.25A$ and $(\mu_L-\mu_R)=0.02A$.

Fig. \ref{spd}(a,b) display the spatial distribution of the different components of the non-equilibrium spin density in FTI and AFTI, respectively. The middle panels show the spatial profile of $S_y$, the spin density component that is aligned along the magnetic order. This component is uniform in FTI and staggered in AFTI, as expected from the magnetic texture of these two systems. We observe finite $S_z$ on both edges, which is a characteristic feature of a TI (bottom panels) and more interestingly finite and oscillatory $S_x$ on both edges (top panels). The oscillation is caused by the scattering at the interfaces between the conductor and leads. Since $S_x$ and $S_y$ are not immune to scalar perturbation, the potential steps at the interfaces mix these components depending on the chirality of each edges. Note that the amplitude of oscillation of $S_x$ in G-AFTI is two orders of magnitude smaller compared to that in FTI which denotes that the scattering in G-AFTI is weaker compared to that in FTI. \par

\begin{figure}[h]
\centering
\includegraphics[width=0.23\textwidth]{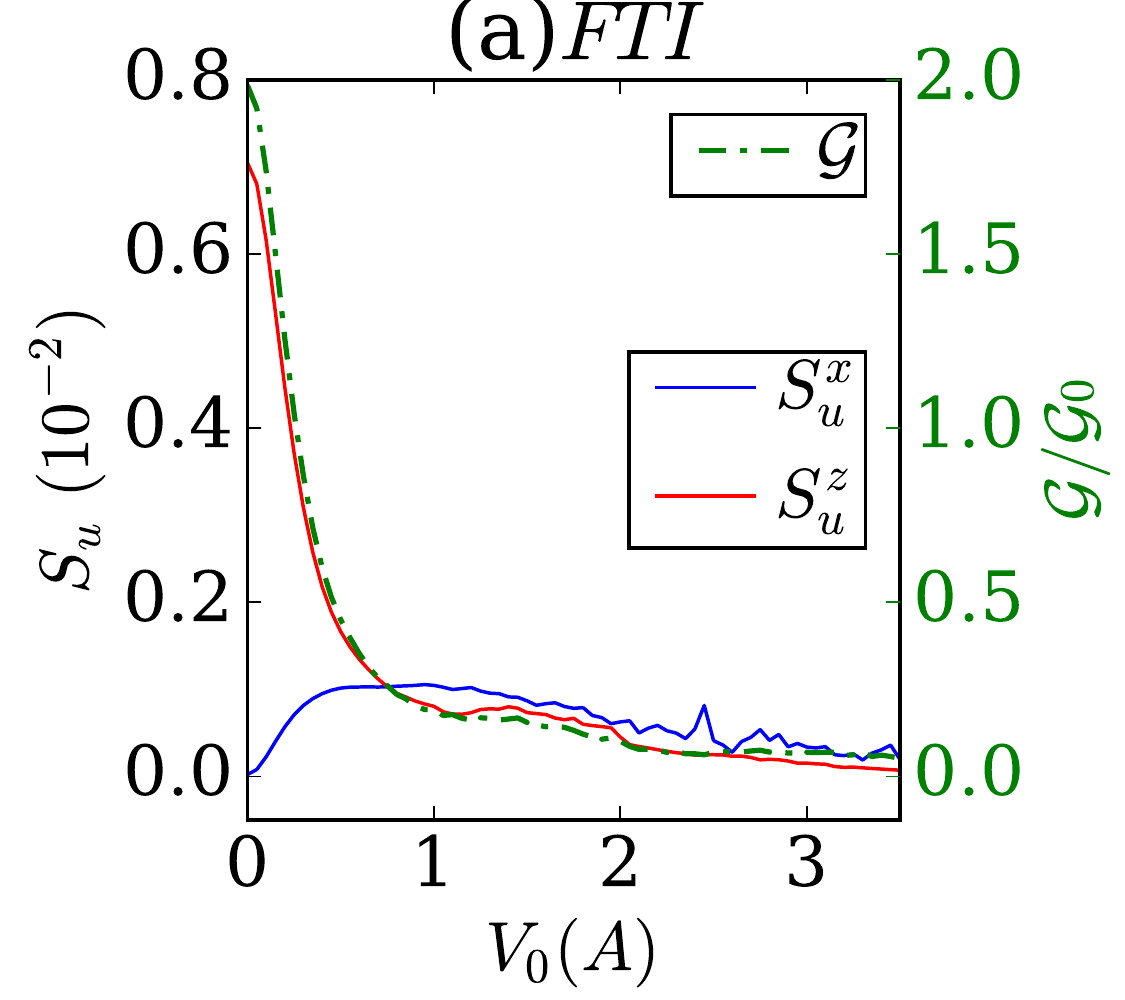}
\includegraphics[width=0.23\textwidth]{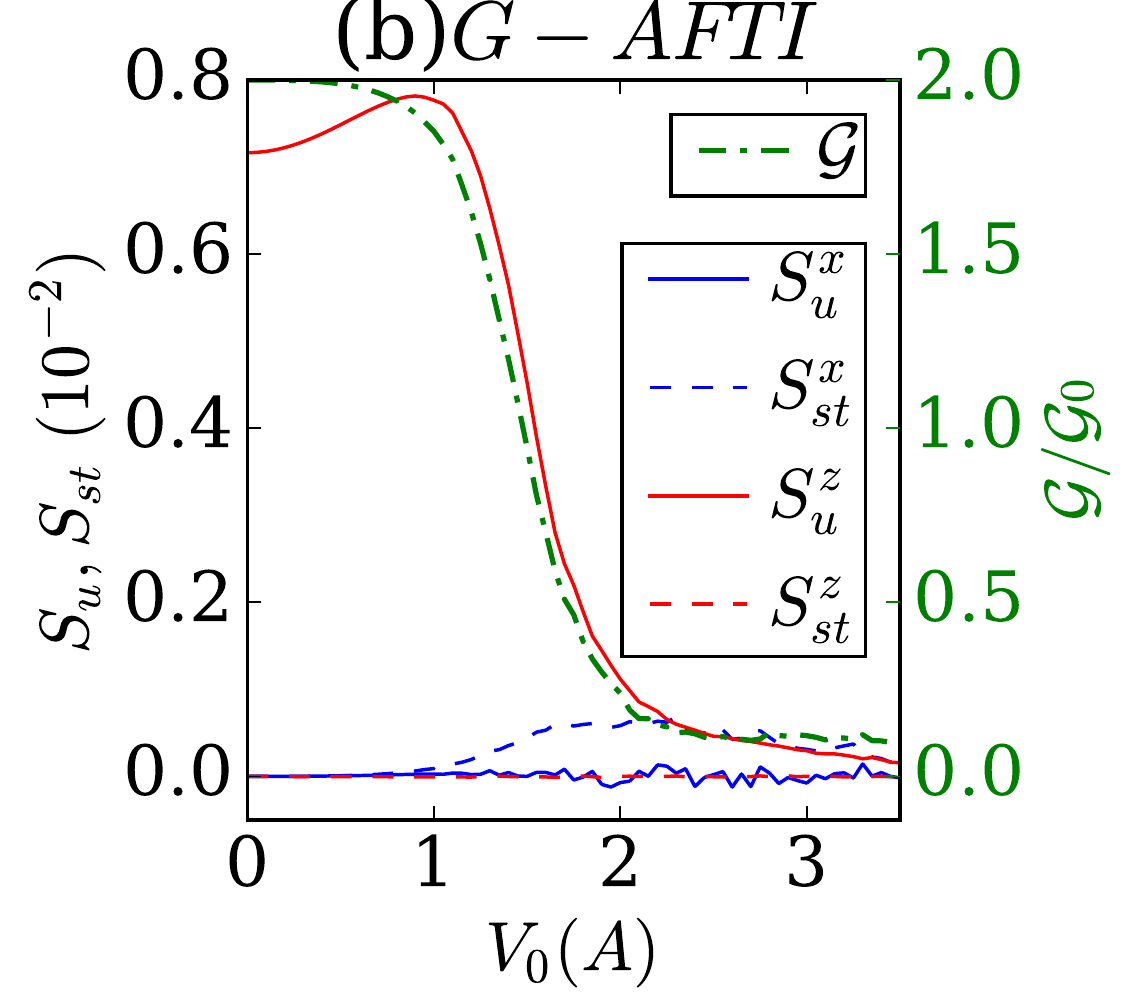}
\caption{Variation of uniform ($S_{\rm u}^{x,z}$) and staggered ($S_{\rm st}^{x,z}$) spin densities as a function of disorder strength with an in plane magnetic order $|\vec{m}_i|=0.2A$ for (a) FTI and (b) G-AFTI. The green dot-dashed line shows the corresponding conductance.}
\label{op}
\end{figure}

From the symmetry we can easily recognize that $S_z$ produces the so-called field-like torque [$\vec{\tau}_F^i \sim \vec{m}_i\times\vec{z}$] and $S_x$ gives rise to the antidamping torque [$\vec{\tau}^i_D \sim \vec{m}_i\times(\vec{z}\times\vec{m}_i)$]. Fig. \ref{spd}(c,d) represent the spatial profile of the field-like ($\tau_{F}$) and antidamping torques ($\tau_{D}$) at the top edges of the FTI and AFTI, respectively. To understand how these torques evolve in the presence of disorder, we further study the robustness of $S_x$ and $S_z$ in presence of scalar disorder (see Fig.~\ref{op}). We define the (i) {\em uniform} spin density ($S^{x,z}_{\rm u} = \langle S^{x,z}_i \rangle$) and (ii) {\em staggered} spin density ($S^{x,z}_{\rm st} = \langle sign(m_i) S^{x,z}_i \rangle$) where the average is over the lattice sites. Since the spin density is localized at the edges we calculate the robustness for the top edge only (Fig.~\ref{op}). Similar results can also be obtained for the bottom edge.

\begin{figure}[h]
\centering
\includegraphics[width=0.46\textwidth]{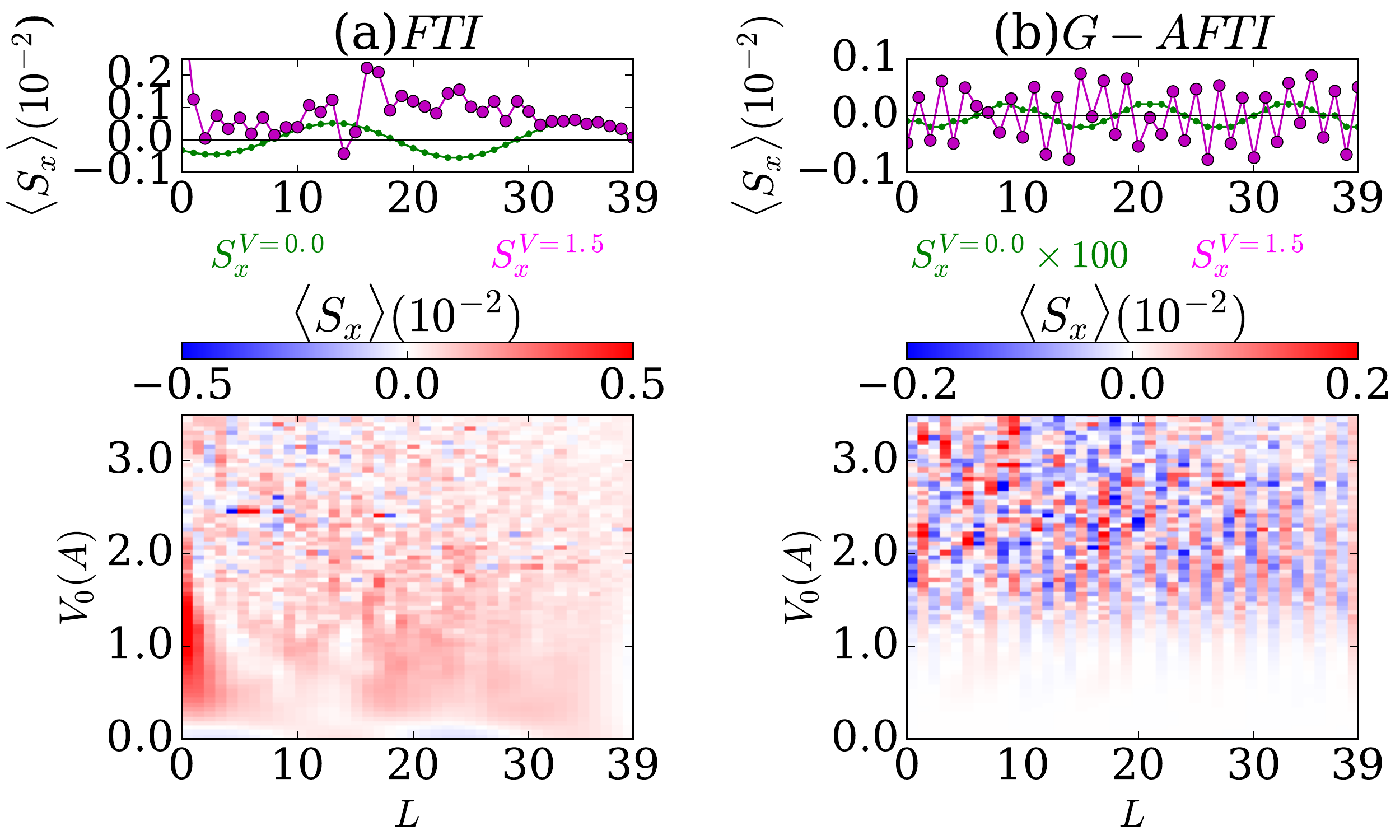}
\caption{Evolution of $S_x$ at top layer with disorder strength for (a)FTI and (b)AFTI. The magnetic order and Fermi level is $|\vec{m}_i|=0.2A$ and $E_F=0.25A$ respectively. The spin density is averaged over 1280 configurations and calculated with a bias voltage $0.02A$.}
\label{vmlr}
\end{figure}

From Fig.~\ref{op} we can see that, correspondingly with conductance, the non-equilibrium spin densities also fall down faster in FTI compared to AFTI. Due to its periodic modulation, $S_{\rm u}^x$ is initially zero for both FTI and AFTI. When increasing the disorder, two different effects take place: (i) a progressive smearing of the edge wave function accompanied by a reduction in $S_{\rm u}^z$; (ii) an increase of disorder-induced spin-dependent scattering resulting in enhanced spin mixing. This disorder-induced spin mixing is at the origin of the $S_{\rm st,u}^x$ component observed in Figs.~\ref{op}(a) and (b). This mechanism has been originally established in metallic spin-valves\cite{Zhang2002} and domain walls \cite{Zhang2004}. In a disordered ferromagnetic device submitted to a non-equilibrium spin density $\vec{S}_0$, spin dephasing and relaxation produce an additional corrective spin density of the form $\sim\vec{m}\times\vec{S}_0$. In the case of FTI, the magnetization is uniform so that a uniform $S_{\rm u}^x \sim S_{\rm u}^z\vec{m}\times\vec{z}$  is produced [Fig.~\ref{op}(a) and Fig.~\ref{vmlr}(a)]. In the case of AFTI, the magnetization is staggered so that a staggered $S_{\rm st}^x \sim S_{\rm u}^z \vec{m}_i\times\vec{z}$  is generated [Fig.~\ref{op}(b) and Fig.~\ref{vmlr}(b)]. In the latter, no uniform $S_{\rm u}^x$ emerges.
Notice that the build-up of $S_{\rm st,u}^x$upon disorder is a non-linear process as disorder increases spin mixing and reduces $S^z_{\rm u}$, at the same time. Hence, one can identify three regimes of disorder. In the case of AFTI displayed in Fig.~\ref{op}(b):
\begin{itemize}
\item From $V_0$=0 to $V_0\approx A$, $S^z_{\rm u}$,  remains (mostly) unaffected, while $S_{\rm st,u}^x$ vanishes on average. 
\item From $V_0\approx A$ to $V_0\approx 2A$, topological protection breaks down progressively and $S^z_{\rm u}$  is reduced upon disorder due to increased delocalization of the edge wavefunction. During this process, disorder enhances spin mixing and thereby $S_{\rm st}^x$ increases moderately. During this moderate increase the reduction of $S^z_{\rm u}$ is compensated by the increase in spin mixing, thereby producing a finite $S_{\rm st}^x$.
\item For $V_0> 2A$, $S^z_{\rm u}$ is further reduced and correspondingly $S_{\rm st}^x$ decreases too, as the disorder-driven spin mixing cannot compensate the reduction of $S^z_{\rm u}$ anymore. 
\end{itemize}
Notice that although $S^z_{\rm u}\gg S_{\rm st}^x$ in the weak disorder limit, both spin density components tend towards a similar value for large disorder ($V_0>2A$), $S^z_{\rm u}\approx S_{\rm st}^x$. In the case of FTI displayed in Fig.~\ref{op}(a), the three regimes appear at different disorder strengths due to the weaker topological protection of the edges states.

To illustrate the progressive build-up of $S^{x}_{\rm st,u}$ upon disorder, the spatial profile of the spin density at the edge of the sample is reported on Fig.~\ref{vmlr} for (a) FTI and (b) AFTI. These calculations confirm that the overall increase in $S^{x}_{\rm st}$ is a direct consequence of spin-dependent scattering upon disorder. In the absence of disorder, the component $S_x$ displays a smooth oscillation (green dots), as discussed above. Following the process described above, in FTI with positive magnetization, the scattering creates mostly positive $S_x$ along the top edge [Fig.~\ref{vmlr}(a)], while in AFTI due to the staggered magnetization $S_x$ acquires a staggered nature [Fig.~\ref{vmlr}(b)].

\begin{figure}[h]
\centering
\includegraphics[width=0.35\textwidth]{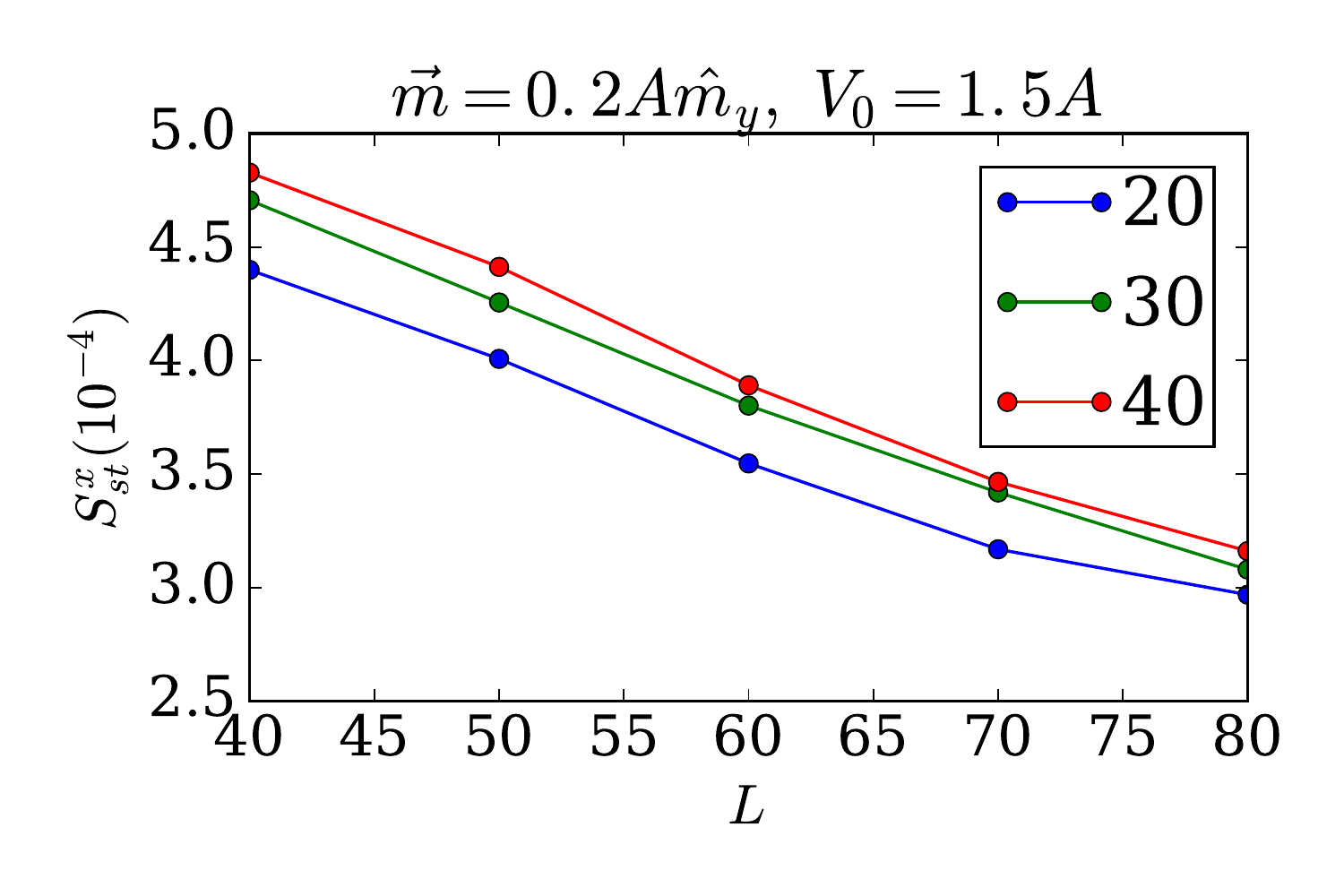}
\caption{Staggered $S^x_{\rm st}$ at $V_0$=1.5A for different length ($L$) and width ($W$, given in legend).}
\label{lwst}
\end{figure}

It is worth mentioning that since the staggered $S_{\rm st}^x$ emerges as a correction to $S_{\rm u}^z$ upon scattering, its magnitude is not only sensitive to disorder but also to the dimension of the channel. As a matter of fact, Fig.~\ref{lwst} shows that for a given amount of disorder, the magnitude of $S_{\rm st}^x$ decreases with increasing the channel length $L$, and (slightly) decreases when reducing the channel width $W$.
We remind that the magnitude of $S_x$ depends on the amplitude of the edge wavefunction as well as on the strength of scattering potential, as mentioned above. Due to finite size effect, the edge localization increases and gradually reaches a saturation value as the width is increased. Therefore for a given disorder strength, the edge states of a wider AFTI undergo smaller delocalization resulting an enhanced $S_{\rm u}^z$ and thereby larger $S_{\rm st}^x$. Increasing the length $L$ favors destructive interferences and therefore reduces $S_{\rm st}^x$ progressively.

It is quite instructive to analyze our results in the light of the latest developments of spin torque studies on antiferromagnets \cite{Gomonay2014, Jungwirth2016, Baltz}. As a matter of fact, it is well known that an external uniform magnetic field only cants antiferromagnetic moments and is unable to switch the direction of antiferromagnetic order parameter. Notwithstanding, {\em time-dependent} uniform magnetic fields (e.g. a magnetic pulse) can induce inertial antiferromagnetic dynamics \cite{Jungwirth2016, Baltz}. In addition, it was recently proposed that a spin torque possessing an antidamping symmetry [i.e. $\vec{m}_i\times(\vec{p}\times\vec{m}_i)$, where $\vec{p}=\vec{z}$ in our case] can manipulate the antiferromagnetic order parameter of a collinear antiferromagnet \cite{Gomonay2010,Zelezny2014}. Applied to the AFTI studied in the present work, these considerations imply that a current pulse can exert a torque on the antiferromagnetic order parameter through the uniform spin density $S^z_{\rm u}$, while a dc current can exert a torque via the staggered spin density $S^x_{\rm st}$. Therefore, the staggered $S^x_{\rm st}$ computed in Figs. \ref{op}(b) and \ref{vmlr}(b) can in principle be used to control the antiferromagnetic order of an AFTI. 

\begin{figure}[h]
\centering
\includegraphics[width=0.155\textwidth]{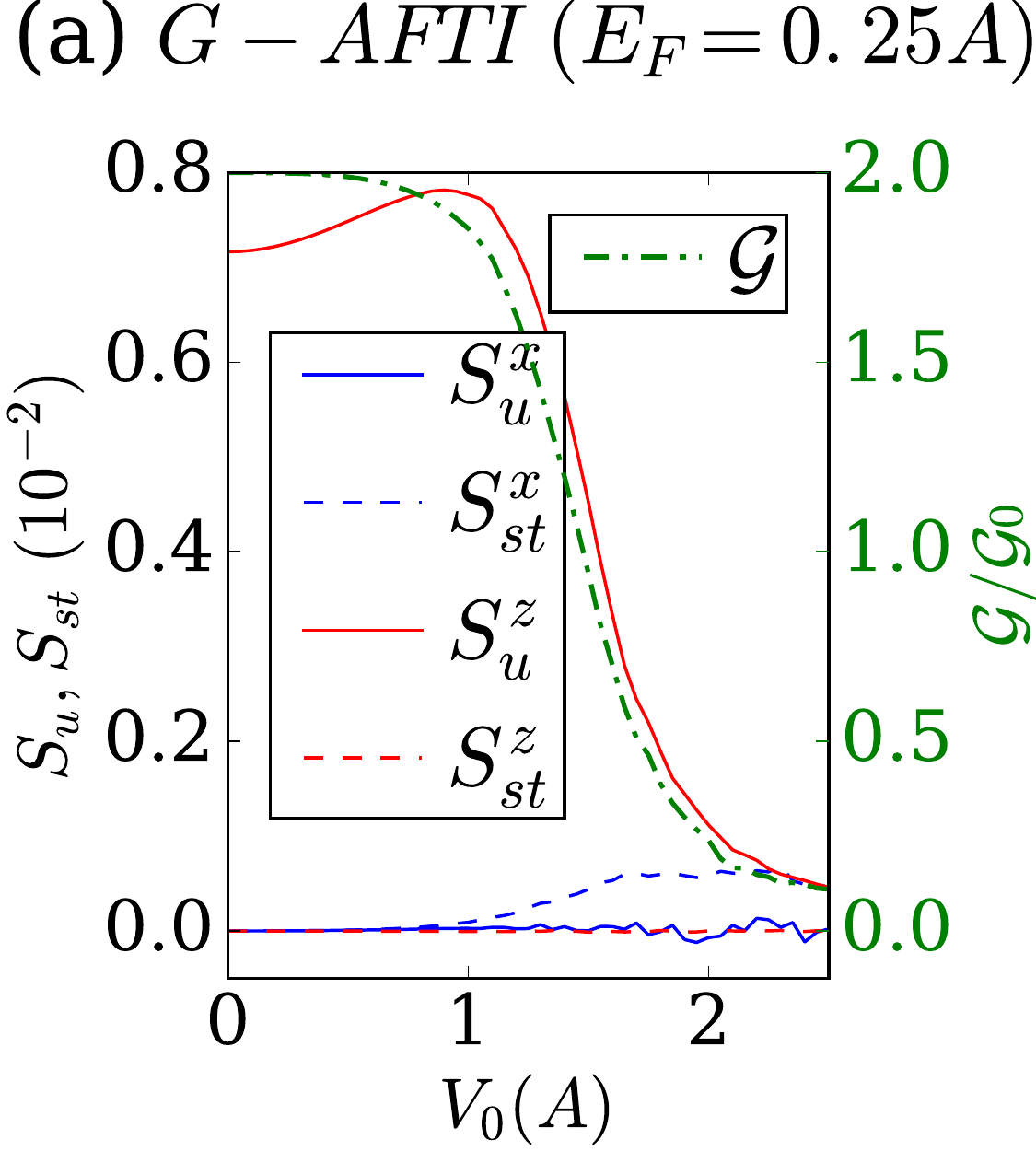}
\includegraphics[width=0.155\textwidth]{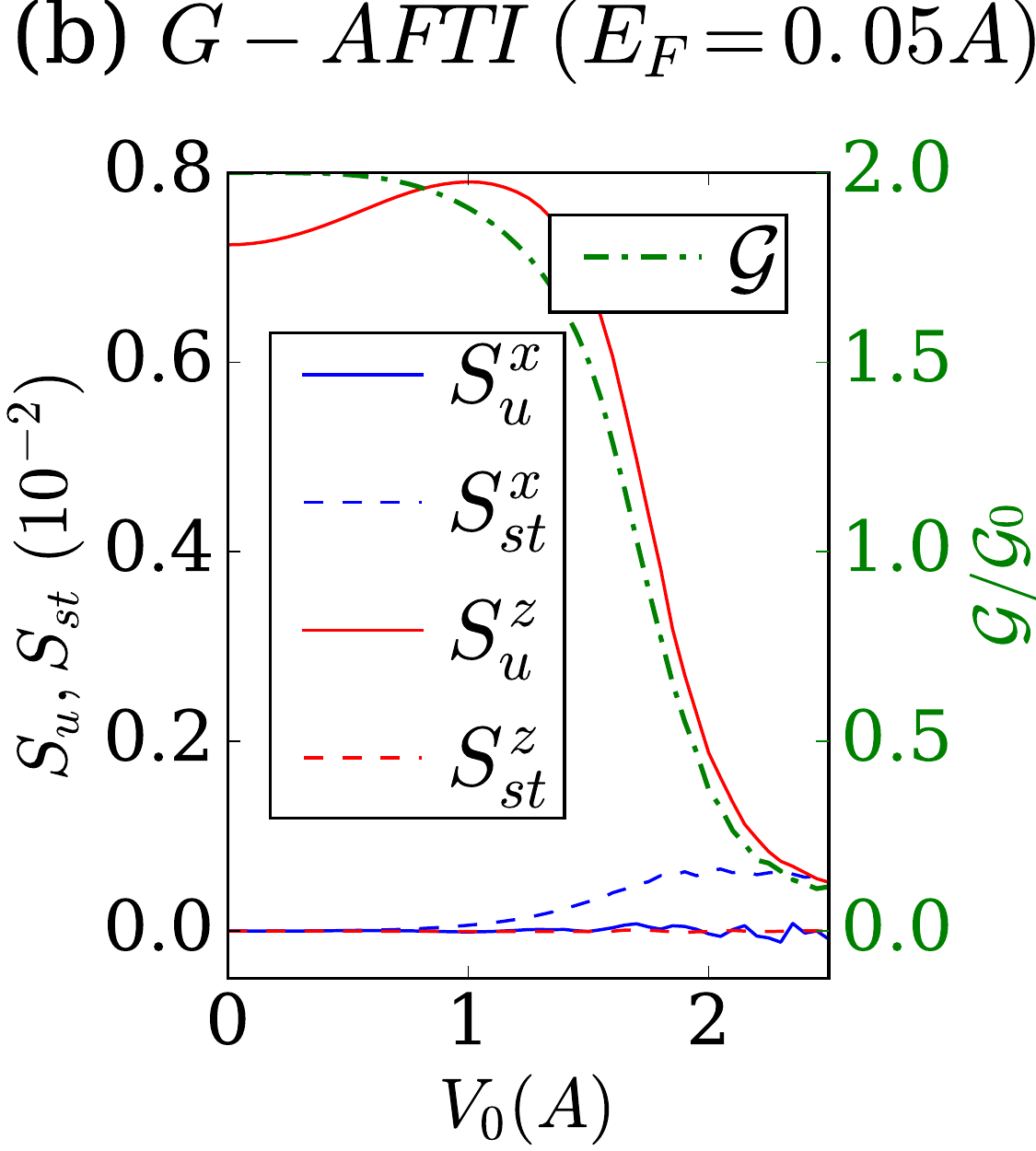}
\includegraphics[width=0.155\textwidth]{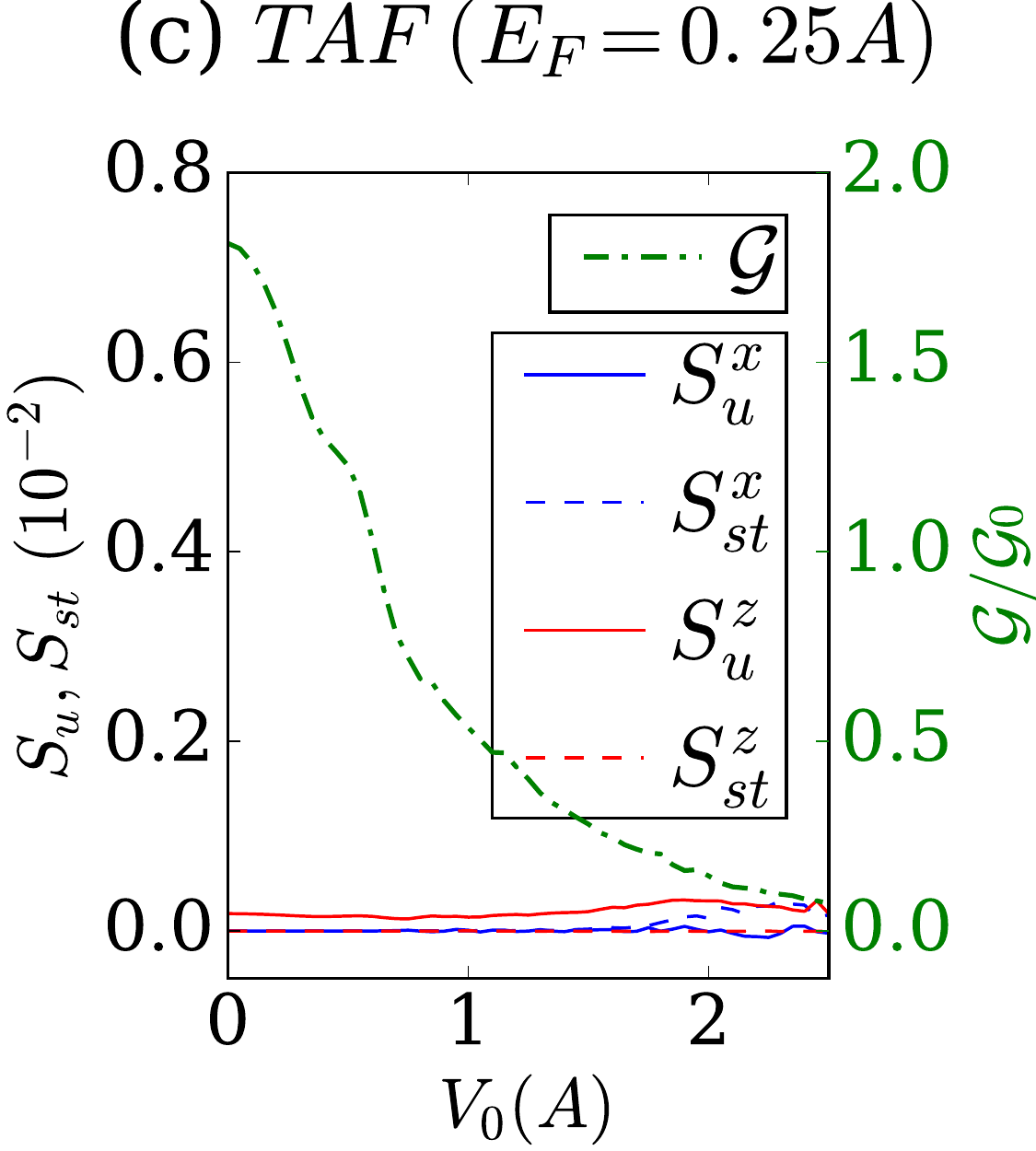}
\caption{The uniform and staggered order parameter for (a) AFTI at $E_F=0.05A$, (b) AFTI at $E_F=0.25A$ and (c) for TAF at $E_F=0.25A$. For TAF we use $B=1.0A, M=-2.2A$. $|\vec{m}_i|=0.2A$ for all three cases.}
\label{stag}
\end{figure}

Note that in case of AFTI, we can choose $E_F$ very close to zero where the topological protection is stronger, without significantly affecting the magnitude of the current-driven spin densities [see Fig.~\ref{stag}(a,b)]. By tuning the parameter $M$ one can weaken the topological protection and turn the AFTI into a trivial antiferromagnet (TAF). In this regime, the non-equilibrium spin density is more distributed within the bulk of the TAF and does not have any topological protection. As a result and in spite of the strong spin-orbit coupling, $S_{\rm u}^z$ and $S_{\rm st}^x$ remain both very small [see Fig. \ref{stag}(c)]. 

\section{Conclusion}
In this work we present a detailed analysis of spin transport in two dimensional FTI and AFTI. We show that topological transport in AFTI is more robust compared to FTI in presence of both out of plane and in plane magnetic order. An in plane magnetic order opens a gap in a FTI but preserves the gapless states in an AFTI when the antiferromagnetic order is along the direction of transport, which allows an AFTI to operate at a much smaller energy. We also study the robustness of the non-equilibrium spin density and SOT against scalar disorder and find that the in-plane spin densities get mixed up due to scattering. In the clean limit, this mixing is two orders of magnitude smaller in AFTI compared to FTI, which suggests that AFTI has stronger topological protection against scalar disorder. The SOT possesses two components, a field-like torque arising from the spin-momentum locking at the edges and an antidamping torque arising from scattering. This antidamping torque linearly decreases when increasing the length of the sample due to destructive interferences.

\section*{Acknowledgement}
This work was supported by the King Abdullah University of Science and Technology
(KAUST) through the Office of Sponsored Research (OSR) [Grant Number OSR-2015-
CRG4-2626].

\bibliographystyle{apsrev4-1}
\bibliography{AFTI1}
\end{document}